\documentclass[aps,prl,showpacs,twocolumn,superscriptaddress]{revtex4}
\usepackage{bm,color}
\usepackage{graphicx}
\usepackage{amsmath, amssymb}
\usepackage{braket}
\usepackage{ulem}
\usepackage[dvipsnames]{xcolor}
\begin{document}
\title{Handwritten Digit Recognition by Spin Waves in a Skyrmion Reservoir}
\author{Mu-Kun Lee}
\affiliation
{Department of Applied Physics, Waseda University, Okubo, Shinjuku-ku, Tokyo 169-8555, Japan}
\author{Masahito Mochizuki}
\affiliation
{Department of Applied Physics, Waseda University, Okubo, Shinjuku-ku, Tokyo 169-8555, Japan}
\begin{abstract}
By performing numerical simulations for the handwritten digit recognition task, we demonstrate that a magnetic skyrmion lattice confined in a thin-plate magnet possesses high capability of reservoir computing. We obtain a high recognition rate of more than 88\%, higher by about 10\% than a baseline taken as the echo state network model. We find that this excellent performance arises from enhanced nonlinearity in the transformation which maps the input data onto an information space with higher dimensions, carried by interferences of spin waves in the skyrmion lattice. Because the skyrmions require only application of static magnetic field instead of nanofabrication for their creation in contrast to other spintronics reservoirs, our result consolidates the high potential of skyrmions for application to reservoir computing devices.
\end{abstract}
\maketitle

\begin{figure}[t]
\centering
\includegraphics[scale=0.4]{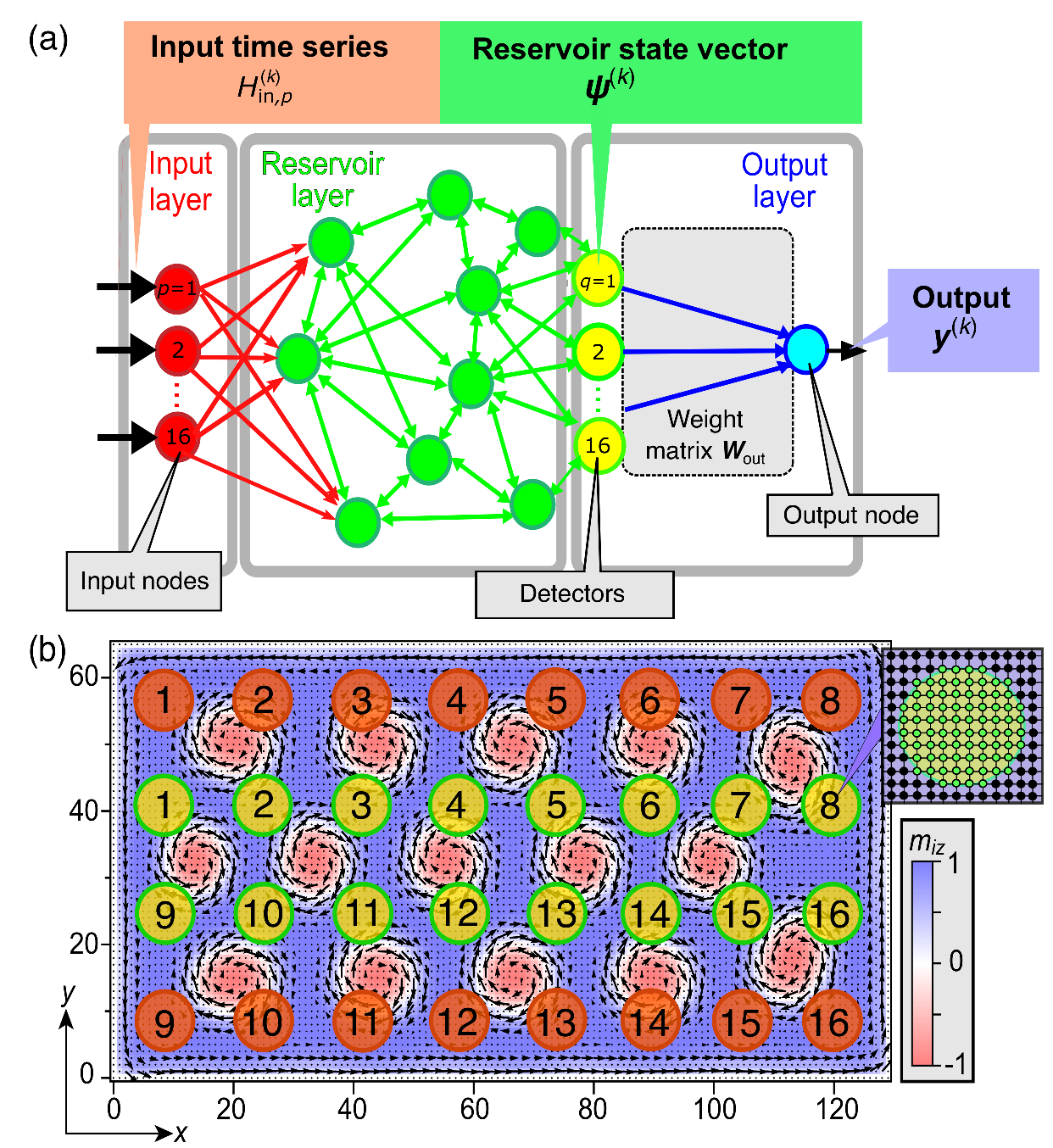}
\caption{(a) Architecture of the reservoir computing. (b) Magnetization configuration of the skyrmion reservoir and the layout of input (red circles) and readout (yellow circles) nodes. In-plane magnetization vectors $(m_{ix}, m_{iy})$ are presented by arrows at sites ($i_x$, $i_y$) when $\text{mod}(i_x, 2)=\text{mod}(i_y, 2)=0$, while out-of-plane components $m_{iz}$ are presented by colors. Inset shows an enlarged view of one readout node.}
\label{Fig01}
\end{figure}
\section{I. Introduction}
Reservoir computing~\cite{Tanaka2019,Nakajima2020,Nakajima2018} is one of the successful derivatives of recurrent neural networks (RNN), which are composed of an input layer, a reservoir, and an output layer [Fig.~\ref{Fig01}(a)]. The reservoir is a dynamical system which maps the input data onto a higher-dimensional information space nonlinearly to mimic the roles of hidden layers in a RNN. In reservoir computing, only the weight matrix $\bm{W}_{\rm out}$ connecting the reservoir and the output nodes is required to be trained, instead of training all the weight matrices linking the layers in a RNN. In this way, the reservoir computing efficiently resolves the difficulties of RNNs regarding the time-consuming and potentially unstable trainings. Aside from purely dynamical reservoir models, many physical reservoirs have been proposed to date~\cite{Optical01,Optical02,Optical03,Optical04,Mechanical01,Mechanical02,Mechanical03,Mechanical04,Biological01,Biological02,Biological03,Biological04,Electronic01,Electronic02,Electronic03,Electronic04,Magnetic01,Kanao2019,STO02,STO03,Furuta2018, Nakane2018,Yamaguchi2020, Bourianoff2018, Prychynenko2018, Pinna2020,Msiska2023,Yokouchi2022,Jiang2019}. Among them, magnetic reservoirs~\cite{Magnetic01,Kanao2019,STO02,STO03,Furuta2018,Nakane2018,Yamaguchi2020}, including the systems involving skyrmions~\cite{Bourianoff2018, Prychynenko2018, Pinna2020,Msiska2023,Yokouchi2022,Jiang2019,Raab}, have several advantages such as nonvolatility~\cite{Grollier2016} (namely, the magnetic system should retain its initial texture after the removal of inputs to fulfill the reproducibility), durability~\cite{YuXZ2011,Zhang2014}, low energy consumption as compared to CMOS architectures~\cite{Joshi2016,Barla,Li2021}, and quick responses~\cite{Kanao2019}.

Magnetic skyrmions are nanometric topological spin textures, which form a hexagonal lattice under the application of a static magnetic field in chiral magnets with broken inversion symmetry~\cite{Muhlbauer2009,YuXZ2010}. Skyrmions are robust against environmental agitations~\cite{YuXZ2011} because of their topological protection~\cite{Nagaosa2013,Braun2012}. A magnetic skyrmion crystal exhibits specific spin-wave modes at microwave frequencies~\cite{Mochizuki2012,Petrova2011,Mochizuki2015}. In this sense, a skyrmion crystal works as a set of series-connected spin-torque oscillators (STOs)~\cite{Magnetic01,Kanao2019,STO02,STO03,Furuta2018} and is expected to possess the characteristics required by a physical reservoir. In contrast to STOs, one advantage of the skyrmion reservoirs is that advanced nanofabrication and complicated manufacturing processes are not required in their production.
In our previous work~\cite{SkRC01}, we have demonstrated that a skyrmion lattice in a thin-plate chiral magnet possesses the fundamental requirements of a reservoir, including the generalization ability, short-term memory, and nonlinearity, inherently carried by the spin wave dynamics in the skyrmion lattice excited by the locally applied magnetic-field pulses as inputs~\cite{Tanaka2019,Nakajima2020,Nakajima2018}.

Skyrmion-based reservoirs have been proposed in literature mostly by random skyrmion configurations both numerically~\cite{Bourianoff2018, Prychynenko2018, Pinna2020,Msiska2023} and experimentally~\cite{Yokouchi2022}. The major purpose of using a random texture is to enhance the nonlinearity of the physical dynamics to increase the potential of separating the linearly inseparable input classes in tasks such as pattern classifications. In this paper, alternatively, we propose to use a (slightly distorted) skyrmion lattice as a reservoir, without introduction of random pinning sites or anisotropies, and we demonstrate that by utilizing the nonlinear interferences of spin waves in such a skyrmion crystal, it suffices to generate high performances in the recognition task of handwritten digits extracted from the Modified National Institute of Standards and Technology (MNIST) database.

For two unrepeated sets of randomly chosen 6600 and 3300 digits as the training and testing datasets, respectively, with 3530 or even smaller amounts of optimized components in the weight matrix, a recognition rate of 88.2\% for the testing dataset can be reached by the skyrmion lattice. This performance is higher than that of 79.3\% by using a dynamical echo-state network model instead of the skyrmion crystal to transform the input data into the reservoir state~\cite{Jaeger01,Jaeger02,Dai} and 50\% by directly using the greyscale data of handwritten digits as state vectors without the operation of any reservoir~\cite{Yokouchi2022}. All training algorithms for these three cases are the same, providing an unbiased comparison of their performances. Importantly, we reveal that this great performance is attributable to the highly nonlinear transformation of input data carried by spin-wave interferences in the skyrmion lattice, instead of merely linear transformations of the input into the amplitudes of spin wave dynamics.

Although the CMOS architecture has been maturely applied to practical handwritten digit recognitions, semiconductor devices are vulnerable to environmental stimuli and cause considerable energy consumption, therefore there is a strong urge to find spintronics alternatives~\cite{Barla,Li2021,Hirohata}. Our result shows that the skyrmion lattice is a promising material for spintronics reservoirs in machine learning applications.

\section{II. Concept and Method}
\subsection{A. Skyrmion spin-wave reservoir}
Figure~\ref{Fig01}(b) shows the design of our skyrmion spin-wave reservoir. Sixteen input nodes (red circles labeled from 1 to 16) are installed near the top and bottom edges of a rectangular skyrmion lattice. Another sixteen nodes (yellow circles labeled from 1 to 16) are installed in between as readout nodes (detectors). We propose in experiments to install circular current loops underlying the nodes to apply and detect local fields via electromagnetic induction. Every node has a radius of 5 lattice constants and contains 80 sites inside. Taking the lattice constant as the length unit, the system size is $128\times 64$. The center of the first input node is located at $(8.5, 55.5)$, and the distances between centers of neighboring nodes are 16 in both of the $x$ and $y$ directions.

\begin{figure*}
	\centering
	\includegraphics[scale=0.76]{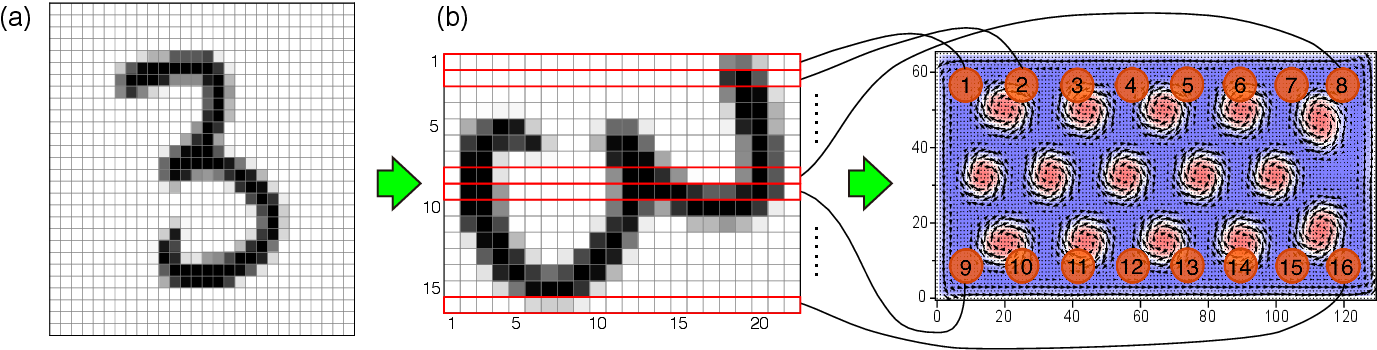}
	\caption{Schematics of the input procedure. (a) One example of the original 28$\times$28 greyscale image extracted from MNIST database. (b) This image is trimmed and rotated by $\pi/2$ to become a 22$\times$16 image. The greyscale data in each of the 16 rows is simultaneously injected into the 16 input areas through magnetic-field pulses, respectively.}
	\label{Fig02}
\end{figure*}
The underlying skyrmion lattice is described by the classical Heisenberg model on a square lattice in the \textit{xy} plane with open boundary conditions. The Hamiltonian for local magnetizations contains the nearest-neighbor ferromagnetic exchange interactions, the Zeeman interactions, and the Dzyaloshinskii-Moriya interactions (DMI) as,
\begin{eqnarray}
\mathcal{H}
&=&-J\sum_{\langle i,j\rangle}\bm m_i \cdot \bm m_j
-\sum_i [\bm H_{\rm ext}+\bm H_{\rm in}(\bm r_i,t)] \cdot \bm m_i
\nonumber\\
&+&D\sum_i (\bm m_i\times \bm m_{i+\hat{\bm x}} \cdot \hat{\bm x} 
+ \bm m_i\times \bm m_{i+\hat{\bm y}} \cdot \hat{\bm y}),
\label{Hamiltonian}
\end{eqnarray}
where $\bm m_i$ denotes the classical unit-length magnetization vector at site $i$. For the Zeeman term, we consider time-dependent local fields $\bm H_{\rm in}(\bm r_i,t)=H_{\rm in}(\bm r_i,t)\hat{\bm z}$, in addition to a static global field $\bm H_{\rm ext}=H_{\rm ext}\hat{\bm z}$ which stabilizes the lattice. The time-dependent fields are applied locally to the sites within the sixteen input-node areas to inject the sequential input data into the skyrmion reservoir. The position vector $\bm r_i=(i_x, i_y)$ of site $i$ represents integer coordinates in units of the lattice constant. We take $J=1$ as the energy unit and set $D=0.36$ and $H_{\rm ext}=0.06$ to produce a stable skyrmion lattice~\cite{SkRC01}.
When $J=1$ meV, the dimensionless time $t=1$ and field strength $H=1$ correspond to 0.66 ps and 8.64 T, respectively.

Magnetization dynamics induced by the applied $\bm H_{\rm in}$ fields are numerically simulated using the fourth-order Runge-Kutta method to solve the Landau-Lifshitz-Gilbert (LLG) equation,
\begin{eqnarray}
\frac{d\bm m_i}{dt}
=\frac{-1}{1+\alpha_{\rm G}^2}\left[
\bm m_i\times \bm H^{\rm eff}_i
+\alpha_{\rm G}\bm m_i \times (\bm m_i \times \bm H^{\rm eff}_i)
\right],
\label{LLGeq}
\end{eqnarray}
where $\alpha_{\rm G}(=0.001)$ is the Gilbert-damping constant, and $\bm H^{\rm eff}_i \equiv -\partial\mathcal{H}/\partial \bm m_i$ is the effective local field. The initial magnetization configuration [Fig.~\ref{Fig01}(b)] is obtained by the Monte Carlo thermalization with simulated annealing to low temperatures followed by a sufficient relaxation executed with the LLG equation.
\subsection{B. Input injection and reservoir state detection}
The procedure of preprocessing and injection of input data begins with the original 28$\times$28 pixel images randomly extracted from the MNIST database with one example as shown in Fig.~\ref{Fig02}(a), each gray-scale pixel representing one of the integers in the range of $[0,255]$. The image is reduced in size to 16$\times$22 by trimming, i,e, three, three, six, and six lines of mostly white pixels are removed from the top, bottom, left, and right edges, respectively. Then each of the trimmed images is rotated by $\pi/2$ as shown in Fig.~\ref{Fig02}(b), and the sixteen rows of greyscales are injected simultaneously into the sixteen input nodes, respectively. Each of the rotated trimmed images is described by a 16$\times$22 input matrix $\bm S_{\rm in}$ whose components $S_{{\rm in},mp}$ are numbers representing the grayscales. In this study, we consider a system with multiple input nodes to reduce the computational cost for the numerical simulations by simultaneously injecting the sixteen sets of sequential pixel data into the skyrmion-lattice reservoir. We note that systems with a fewer input nodes or even with a single input node may also show good performance in the reservoir computing. Such systems might be advantageous in terms of the easy fabrication. Optimization of the number and locations of input and detector nodes to maximize the performance is left for future study.

The 22 components in each of the 16 rows of $\bm S_{\rm in}$ are sequentially entered into the skyrmion reservoir via each of the 16 input nodes as time series of local magnetic-field pulses $H_{\rm in}(\bm r_i,t)\hat{\bm{z}}$. For the input $S_{{\rm in},mp}$ ($1\le m \le 16, 1\le p\le 22$), the magnetic-field pulse is applied to the sites $i$ within an area of the $m$th input node for $(22-p)\Delta t \leq t < (23-p)\Delta t$ to excite the magnetization dynamics. The magnitude of the pulse is set to be $H_{\rm in}(\bm r_i,t)= 5\times10^{-4}S_{{\rm in},mp}$, while the duration of the pulse is fixed at $\Delta t=2.5$. During and after the input injection with these field magnitude and duration, we find the skyrmion lattice retains its overall topological number; namely, the initial number of skyrmions as fifteen in Fig.~\ref{Fig01}(b) remains the same. Therefore, the magnetization excitations belong to the spin wave regime, and the spin waves eventually reach the detectors after experiencing complex reflections and interferences.

To construct the reservoir state of the skyrmion reservoir, during and after the input procedure for each of the $k$th digit, the out-of-plane magnetizations $m_{iz}$ averaged over the sites within each of the $m$th detector area are measured and denoted by $M_{mq}^{(k)}$. Specifically, we record $m_{iz}$ data at instants with a constant interval of $2\Delta t$ for sequentially 22 times (namely, $1\le q \le 22$) starting from the initial time when the first input is injected. Note that the input sequence has a total period of $22\Delta t$ as described in the last paragraph. Therefore, half of the measured $m_{iz}$ data lies in the input procedure, while the other half lies in a period of $22\Delta t$ after the last input is fully injected. In such a way, we expect that the magnetization dynamics dominated by complex spin wave interferences post the input procedure can enhance the nonlinearity and hence the digit recognition rate. (In the end of this paper we design and perform a numerical simulation to consolidate this point.) We then define the reservoir state vector $\bm \psi^{(k)}$ for each of the $k$th input using all elements in the measured signals $\bm{M}^{(k)}$ and an additional constant bias element. The dimension of $\bm \psi^{(k)}$ is therefore 353($=16 \times 22+1$) with 16 detectors each with 22 temporal nodes, plus one constant bias.
\subsection{C. Training and testing procedures}
For the training and testing datasets respectively, we randomly choose unrepeated 6600 and 3300 digits from the MNIST database with equal numbers of digits from 0 to 9 in both sets. Note that we choose a rather small ratio between the numbers of training and testing datasets, namely 2:1 (in literature this ratio is typically 5:1 or higher~\cite{Jiang2019}), as a verification to test the learning ability of the skyrmion lattice. The procedure for the handwritten digit recognition is described below. 

First, we adopt the one-hot representation for the targets of input digits. Specifically, the $k$th target, i.e., the expected output for the $k$th input, is the integer $N^{(k)}$ corresponding to the digit, which is represented by a ten-dimensional vector $\bm x^{(k)}$ where $x^{(k)}_\ell=\delta_{\ell,N^{(k)}+1}\; (\ell=1,2,\cdots,10)$. Here $\delta_{a,b}$ is the Kronecker's delta variable, and $N^{(k)}(=0,1,\cdots,9)$ is the correct number for the $k$th input image. 
Second, since one of the fundamental requirements of the reservoir is its nonlinear mapping of the input data into a higher-dimensional space, we expect all the nonlinear transformations are encoded in the magnetization dynamics in the skyrmion lattice. Therefore, for the output function, only a linear transformation of the reservoir state, $y^{(k)}_\ell = \sum_j W_{\text{out},\ell j} \psi^{(k)}_j$, is executed, with $\bm{W}_{\rm out}$ being a 10$\times$353 output weight matrix. This is distinct from the use of a postprocessing nonlinear softmax output function (such as further defining the output as $y'^{(k)}_\ell = \exp[y^{(k)}_\ell]/\sum_{j=1}^{10} \exp[y^{(k)}_j]$) usually done in RNNs.

Finally, in the training procedure, the components of $\bm{W}_{\rm out}$ are optimized by the gradient descent method with the Adam algorithm~\cite{Adam} to minimize the loss function taken as a mean-square error between the output and target data defined by,
\begin{eqnarray}
L=\frac{1}{N}\sum_{k=1}^{N} \sum_{\ell=1}^{10} (x^{(k)}_\ell- y^{(k)}_\ell)^2.
\label{loss_function}
\end{eqnarray}
Here $N=6600$ is the number of training inputs. In this training procedure, we reset the first and second moments to zero after every 100 iterations in the Adam algorithm defined in~\cite{Adam} to achieve an empirically faster convergence of the loss function. After the training procedure, we calculate $\bm y^{(k)}$ and the recognition rate using the optimized $\bm{W}_{\rm out}$ for both training and testing datasets. The digit of the $k$th image is recognized as the number $\ell-1$ when the $\ell$th component of $\bm y^{(k)}$ is the largest among all of its ten components. The recognition rate is defined as the ratio of the number of correctly recognized images to the total number of images.

\subsection{D. Echo state network}
As a reference for the performance of the skyrmion-lattice reservoir, we use the famous echo state network (ESN) model~\cite{Jaeger01,Jaeger02} to solve the recognition task for the same set of the handwritten digits. For the $k$th digit, the reservoir state $\bm{\psi}^{(k)}_{\rm ESN}$ and output function $\bm{y}^{(k)}_{\rm ESN}$ in an ESN are defined via a nonlinear hyperbolic tangent as
\begin{eqnarray}
\phi^{(k)}_{n,p+1}&=&\text{tanh}\Big[W_{\text{in},nm}S^{(k)}_{\text{in},mp}+W_{\text{res},nm}\phi^{(k)}_{m,p}\Big],\\
\bm{\psi}^{(k)}_{\rm ESN}&\equiv& (\phi^{(k)}_{1,2}, ..., \phi^{(k)}_{1,23},\phi^{(k)}_{2,2}, ..., \phi^{(k)}_{16,2}, ...,\phi^{(k)}_{16,23}, 1)^{\rm T},\nonumber\\
\bm{y}^{(k)}_{\rm ESN}&=&\bm{W}_{\rm out}\bm{\psi}^{(k)}_{\rm ESN}.\nonumber
\end{eqnarray}
Here in the first line the repeated index is summed over, and $\bm{\phi}^{(k)}$ is a 16$\times$23 matrix with the first column being set as $\phi^{(k)}_{m,1}=0$ for all $k,m$~\cite{Dai}. This zero first column is disregarded when constructing the reservoir state as in the second line. Both $\bm{W}_{\rm in}$ and $\bm{W}_{\rm res}$ are fixed 16$\times$16 matrices with elements randomly distributed in $(-1,1)$, while $\bm{W}_{\rm out}$ is a 10$\times$353 output weight matrix to be optimized, whose dimension is the same as that in the skyrmion reservoir described above. To ensure the echo state property, we normalize $\bm{W}_{\rm res}$ by dividing all its elements by $1.01r$, with $r$ being the spectral radius (absolute value of the largest eigenvalue) of the initial unnormalized random matrix~\cite{Furuta2018,Jaeger01,Jaeger02,Dai}, in such a way to make the spectral radius of the normalized matrix become less than one.

As another reference, we consider the digit recognition performed without the reservoir as well. In this case the reservoir state is simply formed by flattening out the 16$\times$22 input matrix $\bm{S}^{(k)}_{\rm in}$ into a 353-component vector with an additional constant bias element. This is purely a linear transformation of the input data. Following the same training procedure described above, the recognition rate is about 50\%. In literature the recognition rates by this linear model range from about 10\% for the handwritten digit recognition~\cite{Yokouchi2022} to 70\% for spoken digit recognition~\cite{Msiska2023}. It may depend on the dimensions of weight matrix, ratio between training and testing data numbers, and/or the specific gradient descent algorithms. Here the same training algorithm as described above is applied for all three performances (i) by the skyrmion lattice, (ii) by the echo state network, and (iii) without reservoir, for an unbiased comparison.
\begin{figure}[tb]
	\centering
	\includegraphics[scale=0.5]{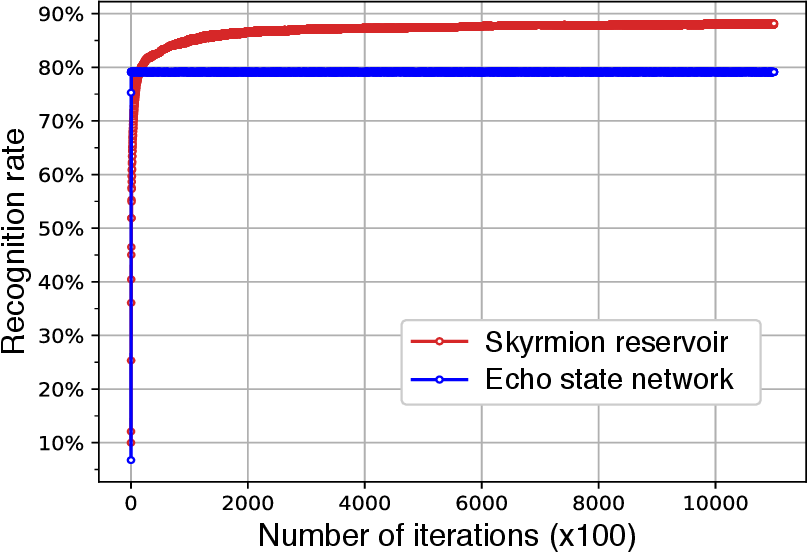}
	\caption{Recognition rate during the training procedure performed by the skyrmion reservoir (red curve) and the echo state network (blue curve).}
	\label{Fig03}
\end{figure}
\begin{figure}[tb]
	\centering
	\includegraphics[scale=0.42]{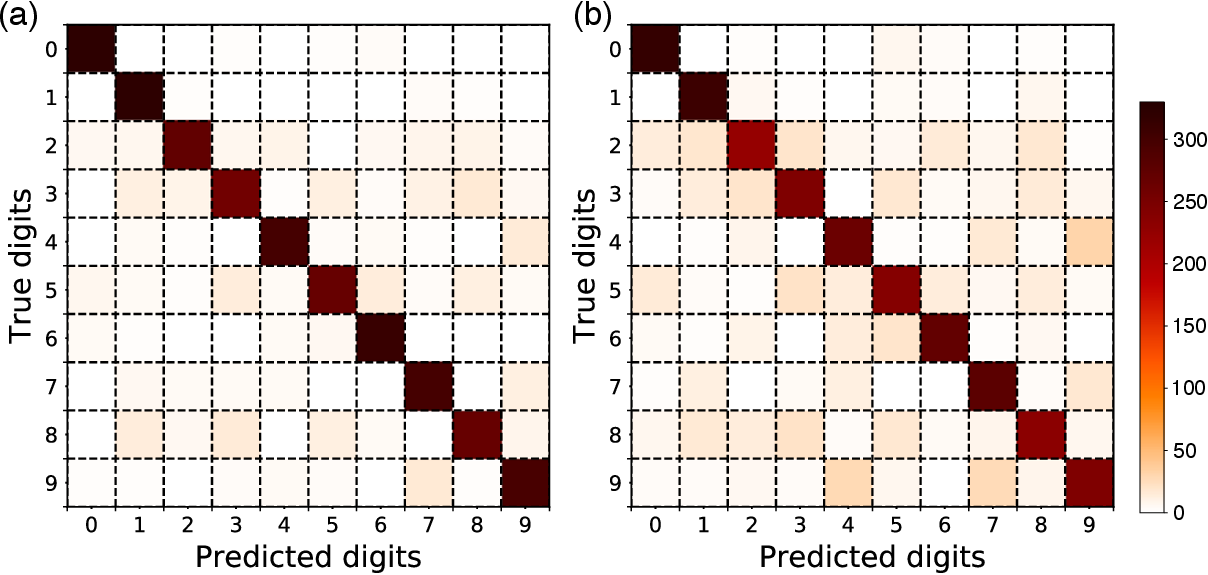}
	\caption{Confusion matrices for (a) skyrmion lattice and (b) echo state network. The diagonal represents the correctly recognized digits. The color intensity reflects the number of recognized digits for each digit class.}
	\label{Fig04}
\end{figure}
\section{III. Result and Discussion}
\subsection{A. Comparison of skyrmion reservoir and echo state network}
In Fig.~\ref{Fig03}, we compare the recognition rates for the testing set performed by the skyrmion reservoir and ESN, plotted as functions of the number of iterations in the training procedure at intervals of 100 iterations. Although the recognition rate by ESN increases faster than that by the skyrmion reservoir, it stops increasing in very initial few hundreds of iterations, and eventually the skyrmion reservoir reaches a recognition rate of 88.18\%, higher by about 10\% than that by ESN (79.3\%). Note that the dimensions of $\bm{W}_{\rm out}$ for both cases are the same, 3530(=$10\times 353$), for a fair comparison of both methods. For a more detailed visualization, in Fig.~\ref{Fig04}(a)-(b) the confusion matrices for both methods are shown, in which the $x$ and $y$ axes represent the predicted and true digits, respectively, and the colormap shows the number of predicted digits in each class. It is apparent that the skyrmion reservoir leads to a better result compared with ESN.

To test the ability of the skyrmion reservoir, we have reduced the temporal sampling of $m_{iz}$ from 22 to 10 for each detectors, leading to a number of 1610($=10\times(16\times10+1)$) weight matrix elements. The recognition rate in this case only slightly drops to $87\%$, still higher than that by ESN with 3530 weight matrix elements. This achievement is comparable to a recent work~\cite{Jiang2019} in which the curren-driven motion of a single skyrmion was utilized for the digit recognition. With 1970 weight parameters, the authors obtained a 87.6\% recognition rate. These facts, together with our previous work~\cite{SkRC01} showing that spin waves in skyrmion lattices are endowed with the input-estimation ability, short-term memory, and nonlinearity, comprehensively reveal that the skyrmion lattice is a suitable candidate for reservoir computing applications.
\begin{figure}[tb]
	\centering
	\includegraphics[scale=0.47]{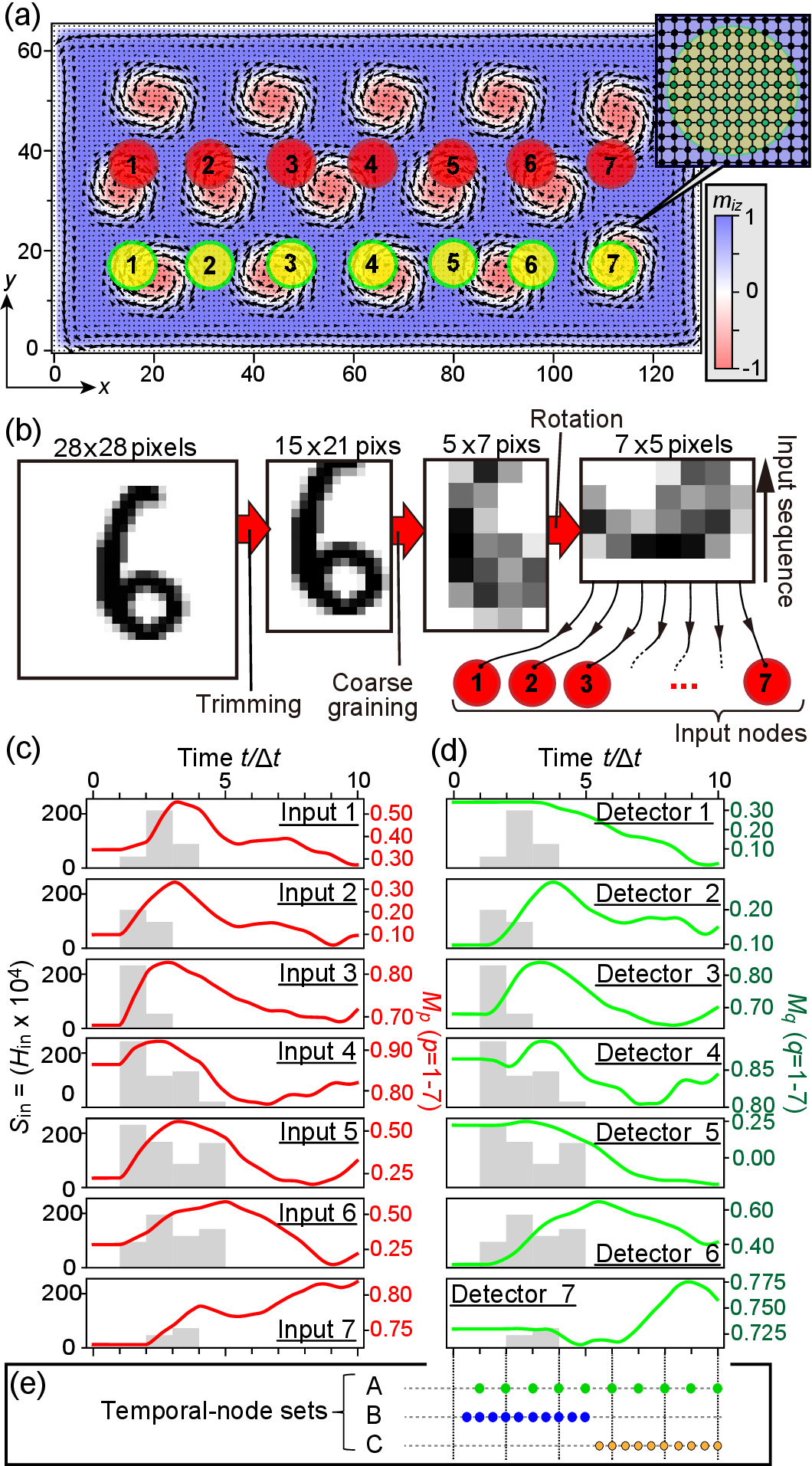}
	\caption{(a) The layout of input (red circles) and detector (yellow circles) nodes for the comparison of linearity and nonlinearity. (b) Schematics of the procedure of input processing and injection. (c) Time profiles of space-averaged out-of-plane magnetizations excited at input nodes 1 to 7. Profiles of input pulses (gray boxes) are also presented. (d) Same as (c) for magnetizations measured at detectors 1 to 7. (e) Three different arrangements of temporal-node sets examined in the comparison. The magnetizations are measured at these moments at each detectors.}
	\label{Fig_grained}
\end{figure}
\begin{figure}[tb]
	\centering
	\includegraphics[scale=0.6]{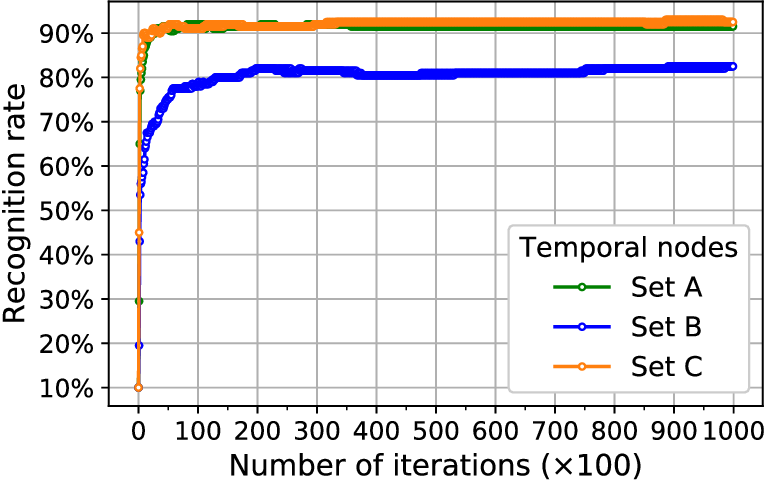}
	\caption{Recognition rate of the testset during the training procedure performed for four three temporal-node sets A-C in Fig.~\ref{Fig_grained}(e).}
	\label{Fig_temp_node_compare}
\end{figure}
\subsection{B. Comparison of linearity and nonlinearity}
In general, the reservoir is required to contain nonlinearity encoded in its states with respect to input data, in order to separate the linearly inseparable classes (digits 0 to 9) of input data. However, the reservoir is often treated as a black box with complex dynamics in which the actual key factors that alter the content of nonlinearity are unclear. In the skyrmion lattice, we find it is possible to compare the performances by two kinds of reservoir states (the $m_{iz}$ data) that are dominated by either a linear or a nonlinear transformation of the input greyscales signals.

For this purpose, we design a numerical test based on a simplified layout with seven input and seven detector nodes as shown in Fig.~\ref{Fig_grained}(a). The center of the first input node is located at $(16.5, 37.5)$, and distance between neighboring nodes is 16 (20) in the $x\ (y)$ direction. To illustrate our idea with a reduced simulation time, we manually choose 400 (200) less distorted digits for the training (testing) dataset, and adopt the preprosseing and input procedure as shown in Figure~\ref{Fig_grained}(b). The digits are trimmed to 15$\times$21 pixels by removing four, three, seven, and six lines of pixels from the top, bottom, left, and right edges, respectively. Then  we coarse grain them into 5$\times$7 pixels as the input matrices $\bm S^{(k)}_{\rm in}$ by taking averages of the greyscales in each 3$\times$3 squares. Each $\bm S^{(k)}_{\rm in}$ is injected into the skyrmion lattice via the seven input nodes as magnetic-field pulses $H_{\rm in}(\bm r_i,t)\hat{\bm{z}}$ with magnitude being $S^{(k)}_{{\rm in},mp} \times 10^{-4}$, while the duration of the pulse is fixed at $\Delta t=10$.
	
Figure~\ref{Fig_grained}(c) shows time profiles of space-averaged out-of-plane magnetization $M_p(t)$ at seven input nodes ($p$=1-7) excited by the applied field pulses for an example digit image of ``6". The figure also shows sequences of the applied five field pulses with gray boxes. The width of box corresponds to the duration of one pulse. We find that the profiles of $M_p(t)$ exhibit the same trend as the instantaneous pulse magnitudes. This is because $M_p(t)$ is directly excited by the applied field pulses at each input node. On the contrary, as seen in Fig.~\ref{Fig_grained}(d), $M_q(t)$ at seven detectors ($q$=1-7) exhibit delayed responses to sequences of the input pulses because the detectors are located away from the input nodes. For detectors 1 and 7, the trends are less correlated, possibly due to edge effects on the spin wave propagations.

Since $M_q(t)$ roughly follows the trends of input streams with a small delay in the early time domain of $0 \leq t \leq 5\Delta t$, it indicates that the magnetization dynamics in this early time domain depends almost linearly on the input signals, because they mostly reach the detectors directly without experiencing significant interferences. On the contrary, $M_q(t)$ in the later time domain of $5\Delta t \leq t \leq 10\Delta t$ is dominantly affected by nonlinear spin wave interferences since the input signals have already stopped. These observations motivate us to compare the performances dominated by linearity and nonlinearity inherent in the reservoir states.

We examine the recognition task for different temporal-node sets at which instants we measure the $m_{iz}$ data for the reservoir state. The sets B and C shown in Fig.~\ref{Fig_grained}(e) contain ten temporal nodes in the early and later time domains, respectively, whose recognition results are plotted in Fig.~\ref{Fig_temp_node_compare} with blue and orange curves. We clearly find that the nonlinearity leads to a better recognition rate (orange curves) by about 10\% than the linearity (blue curves). We note the recognition rate by nonlinearity is even slightly larger than that by temporal-node set A carried out by both linearity and nonlinearity (green curves) with the same number of nodes. This result clearly reveals the power of nonlinearity and, more importantly, shows that its content is potentially adjustable in the spin wave dynamics in skyrmion reservoirs by specialized designs of the input-detector setup, which could be a valuable guideline for future experimental realizations.

For practical applications to general pattern recognitions, we expect that a magnet with a smaller Gilbert damping constant can realize a better performance because the spin waves will have substantially larger amplitudes to enhance the nonlinearity encoded in the magnetization dynamics. Another aspect concerns randomness of the skyrmion lattice. A random skyrmion texture has been proposed as a suitable reservoir by virtue of the large content of nonlinearity~\cite{Bourianoff2018,Pinna2020}. It is generally suggested that when the system is located on the verge of the phase boundary between ordered and chaotic states (so-called edge of chaos), the highest nonlinearity manifests itself~\cite{Chaos01,Chaos02}. These scopes will be our future studies.
\section{IV. Conclusion}
We have theoretically studied the performance of a skyrmion spin-wave reservoir on the handwritten digit recognition task. A high recognition rate of 88.2\% is achievable by the skyrmion reservoir with 3530 parameters in the weight matrix to be trained for a subset of image data randomly extracted from the MNIST database. Importantly, skyrmions emerge spontaneously in magnetic materials with broken inversion symmetry under application of static magnetic field. Therefore, the skyrmion reservoir requires no advanced nano-fabrications for production, in contrast to other spintronics reservoirs using, e.g., spin-torque oscillators and magnetic tunnel junctions~\cite{Magnetic01,Kanao2019,STO02,STO03,Furuta2018,Nakane2018}. Recently, even a zero-field skyrmion lattice is possible to be stabilized experimentally~\cite{Zerofield01,Zerofield02,Zerofield03}, raising up more possibilities for its spintronics applications. Our work paves the way to realizing high-performance spintronics reservoirs in machine-learning applications.

This work is supported by Japan Society for the Promotion of Science KAKENHI (Grant No. 20H00337), CREST, the Japan Science and Technology Agency (Grant No. JPMJCR20T1), and a Research Grant in the Natural Sciences from the Mitsubishi Foundation.

\end{document}